\pdfoutput=1
\documentclass[10pt]{iopart}
\usepackage{graphicx}
\usepackage{latexsym}
\usepackage{url}
\usepackage{xcolor}

\def\sun{\odot}
\def\h0units{\mathrm{km\,s^{-1}\,Mpc^{-1}}}
\def\cunits{\mathrm{km\,s^{-1}}}
\newcommand{\om}{\Omega_{\rm M}}
\newcommand{\ok}{\Omega_K}
\newcommand{\ola}{\Omega_{\Lambda}}

\def\aap{A\&A\,  }

\def\aj{AJ  }
\def\apj{ApJ\,  }
\def\apjs{ApJS  }

\def\jcap{Journal of Cosmology and Astroparticle Physic  } 
 
\def\mnras{MNRAS\,  }

\begin{document}
\title
{
The Great Wall of SDSS galaxies 
}
\vskip  1cm
\author     {Lorenzo  Zaninetti}
\address    {Physics Department,
 via P.Giuria 1,\\ I-10125 Turin,Italy }
\ead {zaninetti@ph.unito.it}
\begin {abstract}
An enhancement  in the number of galaxies
as function of the redshift 
is visible on the SDSS Photometric Catalogue DR 12
at z=0.383. 
This over-density  of galaxies 
is named the Great Wall.
This variable number of galaxies as a function of the 
redshift can be explained  in the framework of
the luminosity function for galaxies.
The differential  of the luminosity distance 
in respect to the redshift is evaluated 
in the framework of the LCDM cosmology.
\end{abstract}
\vspace{2pc}

\maketitle

\section{Introduction}

We review some early works
on the  "
CfA2
  Great Wall", which is
the name  that was introduced by  \cite{Geller1989}
to classify an enhancement in the number of galaxies 
as a function of the redshift
that is visible in the Center for Astrophysics (CfA) redshift survey  
\cite{deLapparent1989}.
The evaluation of the two point correlation function
was done by  \cite{Ramella1992} on the three slices of the CfA redshift
survey.
A careful analysis was performed on Sloan Digital Sky Survey (SDSS) 
DR4 galaxies by
\cite{Deng2007} : 
the great wall 
was detected in the range  $ 0.07 < z< 0.09$.
The substructures, the morphology  and the galaxy contents were analyzed  by 
\cite{Einasto2010,Einasto2011} and 
the luminosities and  masses 
of galaxies were discussed
in the framework of SDSS-III's Baryon Oscillation Spectroscopic Survey (BOSS)
\cite{Einasto2017}.
The theoretical explanations for the Great Wall
include an analysis 
of 
the peculiar velocities, see 
\cite{dellAntonio1996a,dellAntonio1996b},
the nonlinear fields 
of the Zel'dovich approximation,
see  \cite{Zeldovich1970,Shandarin2009,Shandarin2011}, 
and the cosmology of the Great Attractor, see 
\cite{Valkenburg2012}. 
The layout of the rest of this paper is as follows.
In Section \ref{preliminaries}, 
we introduce the LCDM cosmology and the 
luminosity function for galaxies.
In Section \ref{photomaximum}, 
we introduce the adopted catalog for galaxies 
and the theoretical  basis of the maximum 
for galaxies as function of the redshift.

\section{Preliminaries}
\label{preliminaries}
This section introduces an approximate  luminosity 
distance as a function of the redshift  in LCDM and derives 
the connected differential.
The Schechter luminosity function for galaxies is reviewed.

\subsection{Adopted cosmology}

Some useful formulae  in $\Lambda$CDM cosmology
can be expressed in terms of a Pad\'e approximant.
The  basic parameters 
are: 
the Hubble constant, $H_0$, expressed in  $\h0units$,
the velocity of light, $c$,  expressed in $\cunits$, and
the three numbers $\om$, $\ok$, and $\ola$,
see \cite{Zaninetti2016a} for more details.
In the case of  the Union 2.1 compilation, see
\cite{Suzuki2012},
the parameters are  $H_0 = 69.81 \h0units$, $\om=0.239$  and  $\ola=0.651$.
To have the  luminosity distance, $D_L(z;H_0,c,\om,\ola)$, 
as a  function of the redshift only, 
we apply     the minimax rational approximation,
which is characterized by the two parameters $p$ and $q$.
We  find    a simplified expression
for the luminosity distance, $D_{L,6,2}$,
when $p=6$ and $q=2$ 
\begin{eqnarray}
D_{L,6,2}= 
\frac
{
ND
}
{
0.284483+ 0.153266\,z+ 0.0681615\,{z}^{2}
} 
\\ 
\quad  for \quad 0.001 <z<4
\quad ,
\nonumber
\label{dlz}
\end {eqnarray}
where 
\begin{eqnarray}
ND = - 0.0017+ 1221.80\,z+ 1592.35\,{z}^{2}+ 504.386\,{z}^{
3}
\nonumber \\
+ 85.8574\,{z}^{4}+ 0.41684\,{z}^{5}+ 0.186189\,{z}^{6}
\quad ,
\end{eqnarray}

The inverse of the above function, i.e. the redshift $z_{6,2}$
as function of the luminosity distance, is 
\begin{eqnarray}
z_{6,2} =
3.3754\,10^{-5}D_{{L}}- 0.46438
 +{ 2.1625\times 10^{-14}} \times
\nonumber  \\
\,\sqrt { 2.4363\,10^{18}{D_{{L}}}^{2}+{ 3.9538\times 10^
{23}}\,D_{{L}}+{ 4.6114\times 10^{26}}}
\quad .
\label{zinverse}
\end{eqnarray}

\subsection{Luminosity function for galaxies}

We used the   Schechter function, 
see  \cite{schechter},
as a luminosity function (LF) for 
galaxies
\begin{equation}
\Phi (L) dL  = (\frac {\Phi^*}{L^*}) (\frac {L}{L^*})^{\alpha}
\exp \bigl ( {-  \frac {L}{L^*}} \bigr ) dL \quad  ,
\label{equation_schechter}
\end {equation}
here $\alpha$ sets the slope for low values
of luminosity, $L$,  $L^*$ is the
characteristic luminosity and $\Phi^*$ is the normalisation.
The equivalent distribution in absolute magnitude is
\begin{equation}
\Phi (M)dM=0.921 \Phi^* 10^{0.4(\alpha +1 ) (M^*-M)}
\exp \bigl ({- 10^{0.4(M^*-M)}} \bigr)  dM \quad ,
\label{lfstandard}
\end {equation}
where $M^*$ is the characteristic magnitude as derived from the
data.
The scaling with  $h$ is  $M^* - 5\log_{10}h$ and
$\Phi^* ~h^3~[Mpc^{-3}]$.

\section{The photometric maximum}
\label{photomaximum}

This section models   
the Great Wall that is visible on the SDSS Photometric Catalogue DR 12.
It also evaluates the theoretical  number of galaxies as a
function of the redshift.

\subsection{The SDSS data}

\begin{figure}
\begin{center}
\includegraphics[width=10cm]{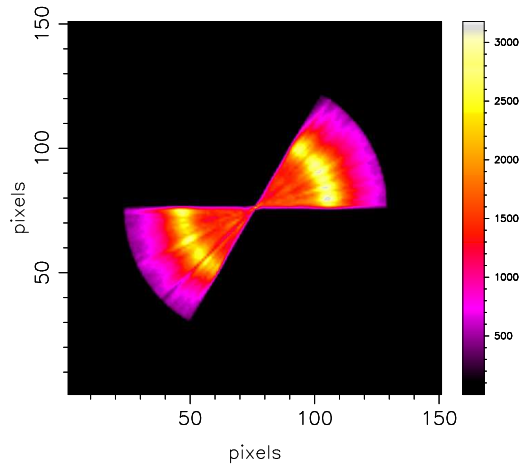}
\end{center}
\caption
{
Contour for the number of galaxies (u-band)  for areas 
of $0.025\times0.025$.
}
\label{wall_heat}
\end{figure}

\begin{figure}
\begin{center}
\includegraphics[width=10cm]{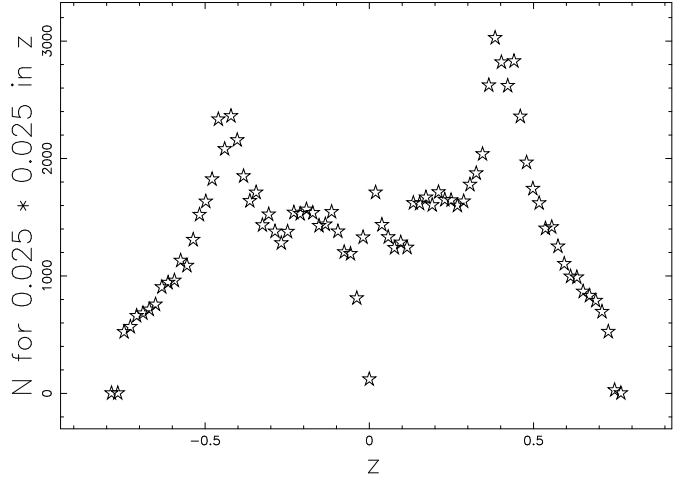}
\end{center}
\caption
{
The number of galaxies (u-band) along a line for areas 
of $0.025\times0.025$.
}
\label{wall_linea}
\end{figure}

We processed the  SDSS Photometric Catalogue DR 12,
see \cite{Alam2015}, which contains  10450256 galaxies 
(elliptical +
spiral)
with redshift.
In the following  we will use the generic
term galaxies without distinction between the  two types,
elliptical and
spiral.
The number of galaxies for an area in redshift  
of  $0.025\times0.025$ of the u-band  is reported 
in Figure  \ref{wall_heat} as a contour plot
and in Figure \ref{wall_linea} as a cut along a line.

\subsection{The theory}

The flux, $f$,
is
\begin{equation}
f  = \frac{L}{4 \pi r^2}
\quad ,
\label{flux}
\end{equation}
where $r$ is the luminosity  distance.
The luminosity distance is  
\begin{equation} 
r=D_{L,6,2}
\quad ,
\end{equation}
and  the relationship between $dr$
 and $dz$ is
\begin{equation}
dr =  \frac
{
N
}
{
D
}
dz
\quad  ,
\end{equation}
where
\begin{eqnarray}
N=74813.67+ 10.9263\,{z}^{7}+ 49.05768\,{z}^{
6}+ 2642.64259\,{z}^{5}
\nonumber \\
+ 16024.51314\,{z}^{4}+
 54307.16663\,{z}^{3}
+ 127258.486\,{z}^{2}+
\nonumber  \\
 195005.8564\,z
\quad ,
\end{eqnarray}
and
\begin{equation}
D= \left( {z}^{2}+ 2.248575472\,z+ 4.173664398 \right) ^{2}
\quad .
\end{equation}
The joint distribution in {\it z}
and  {\it f}  for the number of galaxies
 is
\begin{equation}
\frac{dN}{d\Omega dz df} =
\frac{1}{4\pi}\int_0^{\infty} 4 \pi r^2 dr
\Phi (L;L^*,\sigma)
\delta\bigl(z- (z_{6,2})\bigr)
\delta\bigl(f-\frac{L}{4 \pi r^2}    \bigr)
\quad ,
\label{nfunctionzsch}
\end{equation}
where $\delta$ is the Dirac delta function,
$\Phi (L;L^*,\sigma)$  has  been defined
in equation~(\ref{equation_schechter})
and $z_{6,2}$ has  been defined  in 
equation (\ref{zinverse}).
The explicit version  is
\begin{equation}
\frac{dN}{d\Omega dz df} =\frac{NN}{DD}
\quad ,
\label{nzwall}
\end{equation}
where 
\begin{eqnarray}
NN= 
644.3\, \left( z+ 1.058 \right) ^{4} \left( z-
 0.000001412 \right) ^{4} \times
\nonumber  \\
\left( {z}^{2}+
 4.889\,z+ 13.34 \right) ^{4} 
\left( {z}^{
2}- 3.708\,z+ 464.6 \right) ^{4} \times
\nonumber \\
{{\rm e}^{
{\frac {{\it AA}}{{{\it BB}}^{2}{\it L^*}}}}} \left( {\frac 
{
CC
}
{{{\it BB
}}^{2}{\it L^*}}} \right) ^{\alpha}\times
\nonumber \\
{\it \Phi^*}\, \left( z+
 0.5328 \right)  \left( {z}^{2}+ 5.047\,z+
 7.9141 \right)
\nonumber  \\
  \left( {z}^{2}+ 0.7\,z+
 7.011 \right)  \left( {z}^{2}- 1.79\,z+
 231.58 \right) 
\quad ,
\end{eqnarray}
\begin{eqnarray}
CC=f
 \left( z+ 1.058 \right) ^{2} \left( z-
 0.0000014124 \right) ^{2} \times
\nonumber \\
\left( {z}^{2}+
 4.889\,z+ 13.343 \right) ^{2} \left( {z}^{
2}- 3.708\,z+ 464.6 \right) ^{2}
\end{eqnarray}
 
\begin{equation}
DD = \left( {z}^{2}+ 2.248\,z+ 4.173 \right) ^{6}{\it L^*}
\end{equation}
\begin{eqnarray}
AA=- 93.76\,f{z}^{12}- 419.8\,f{z}^{11}- 86945.4\,f{z}^{
10}
\nonumber \\
- 701622.4\,f{z}^{9}- 22679411\,f{z}^{8}- 239083307\,f{z}^{
7}
\nonumber \\
- 1430432291\,f{z}^{6}- 4912205831\,f{z}^{5}
\nonumber  \\
+ \left( 
 4.540788\,\alpha\,{\it L^*}- 10191896880\,f \right) {z}^{4}+
\nonumber \\
 \left(  20.4206\,\alpha\,{\it L^*}- 10524532830\,f \right) {z
}^{3}
+ \left(  60.862\,\alpha\,{\it L^*}- 4037704632\,f
 \right) {z}^{2}
\nonumber \\
+ \left(  85.22 \,\alpha\,{\it L^*}
+
 11406.2 \,f \right) z+ 79.098 \,\alpha\,{\it L^*}-
 0.008055\,f
\quad , 
\end{eqnarray}
\begin{equation}
BB={z}^{2}+ 2.2485\,z+ 4.173
\quad .
\end{equation}

\begin{figure}
\begin{center}
\includegraphics[width=6cm]{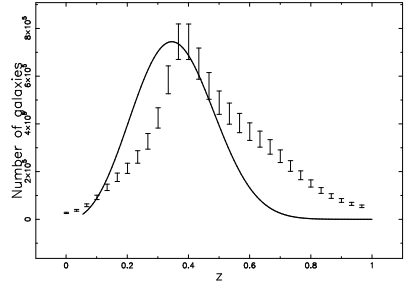}
\end{center}
\caption
{
The galaxies of SDSS DR 12 (u-band)  with   
$ 400
 \, L_{\sun}/Mpc^2  \leq
f \leq  5 \,10^7  \, L_{\sun}/Mpc^2  $
are  organized by frequency versus
distance (empty circles) and
the error bar is given by the square root of the frequency.
The maximum frequency of the observed galaxies is
at    $z=0.38$.
The full line is the theoretical curve that is
generated by
$\frac{dN}{d\Omega dr df}$
as given by the application of the Schechter  LF,
which  is Equation~(\ref{nzwall}),
and the theoretical  maximum is at 
$z=0.382 $.
The parameters are
$L^*= 2.038 \,10^{10} L_{\sun}$   and
$\alpha$ =-0.9 \,.
}
          \label{wall_maximum}%
    \end{figure}

\begin{figure}
\begin{center}
\includegraphics[width=10cm]{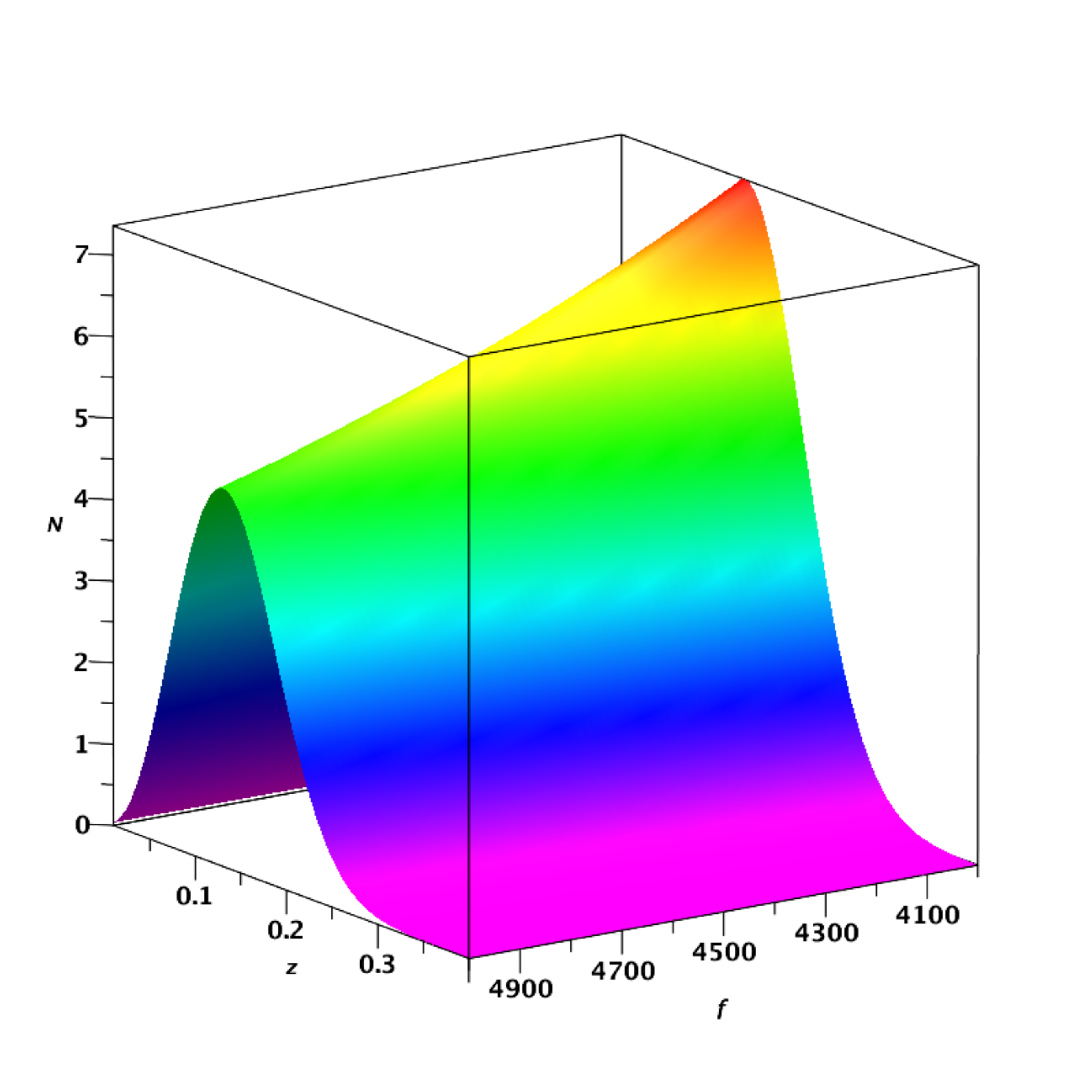}
\end{center}
\caption
{
The theoretical number of galaxies divided by 1000 
as a function of redshift and  flux is expressed
in  $L_{\sun}/Mpc^2$.
The parameters are
$L^*= 2.038 \,10^{10} L_{\sun}$,
$\alpha$ =-0.9
and
$\frac {\Phi^*}{Mpc^{-3}}=0.038$ \,.
}
          \label{wall_fluxz}%
    \end{figure}

\begin{figure}
\begin{center}
\includegraphics[width=10cm]{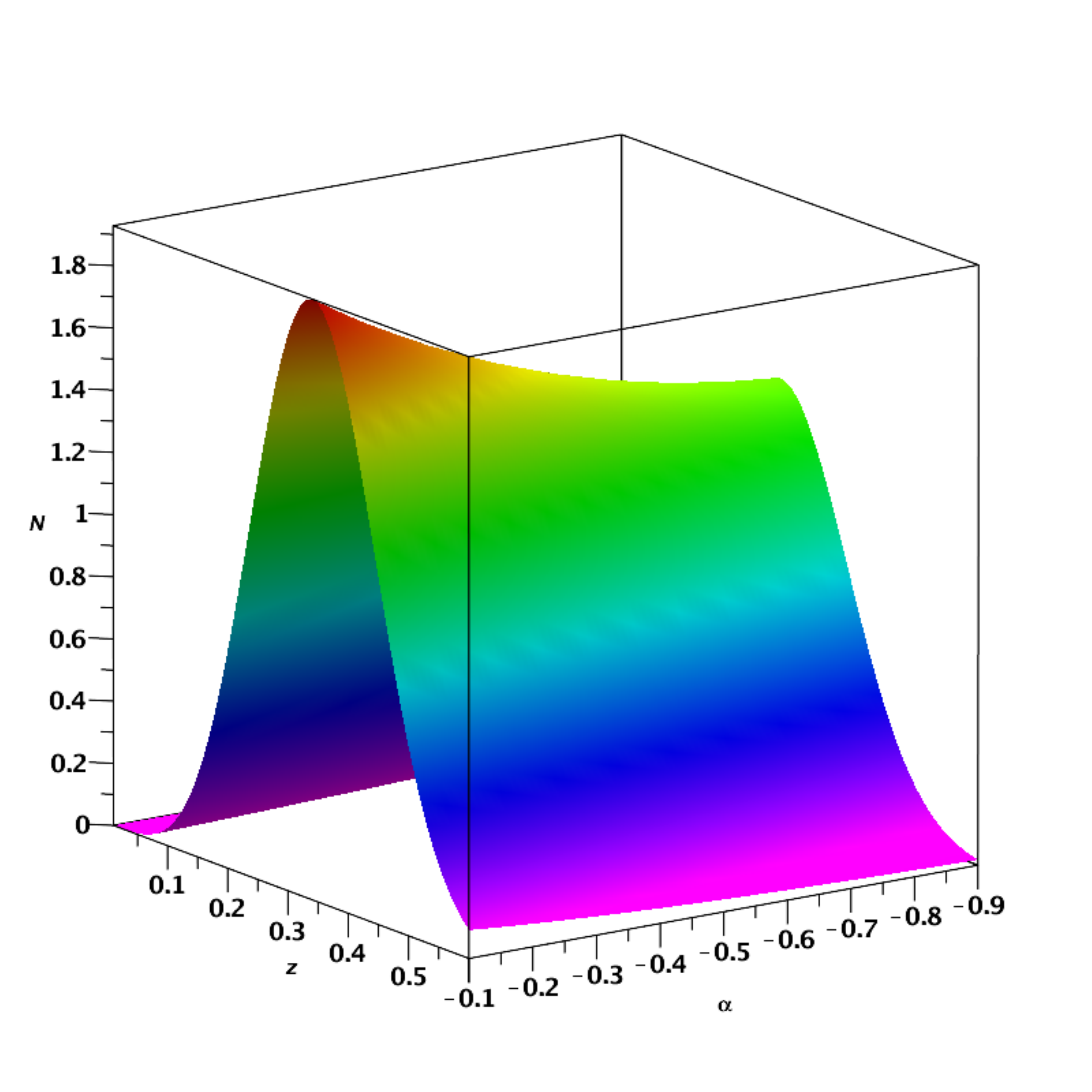}
\end{center}
\caption
{
The theoretical number of galaxies divided by 100000
as a function of $\alpha$ and redshift
when
$L^*= 2.038 \,10^{10} L_{\sun}$,
$f=1000\,L_{\sun}/Mpc^2$, 
and
$\frac {\Phi^*}{Mpc^{-3}}=0.038$ \,.
}
          \label{wall_alpha}%
    \end{figure}

Figure~\ref{wall_maximum}
presents the number of  galaxies  that are observed in 
SDSS DR 12
as a function  of the redshift  for  a given
window in  flux, in addition to the theoretical curve.
The theoretical number of galaxies
is reported in Figure~\ref{wall_fluxz}
as a function of the flux  and redshift,
and is reported in Figure~\ref{wall_alpha}
as a function of $\alpha$ and redshift.

\begin{figure}
\begin{center}
\includegraphics[width=6cm]{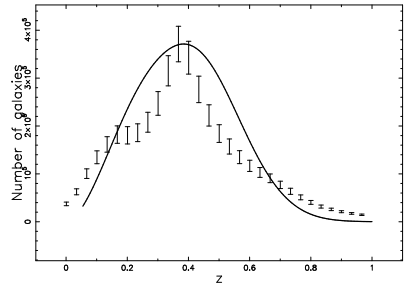}
\end {center}
\caption
{
All of the  galaxies  of the SDSS DR12  catalog (u-band) are 
organized in frequencies versus
spectroscopic   redshift.
The error bar is given by the square root of the frequency
(Poisson
distribution).
The maximum frequency of all observed galaxies is
at  $z=0.383$.
The full line is the theoretical curve
generated by
$\frac{dN}{d\Omega dz}(z)$
as given by the numerical integration
of  Eq.~(\ref{integrale_wall_tutte}) with
$L^*= 2.038 \,10^{10} L_{\sun}$,
$\alpha =-0.9$ 
and
$\frac {\Phi^*}{Mpc^{-3}}=0.038$ \,.
}
          \label{wall_all}%
    \end{figure}

The total number of galaxies
comprised between a minimum value of flux,
$f_{min}$, and a maximum value of flux $f_{max}$,
for the Schechter LF
can be computed through the integral
\begin{equation}
\frac{dN}{d\Omega dz} = \int_{f_{min}} ^{f_{max}}
\frac{NN}{DD}
\,df
\quad.
\label{integrale_wall_tutte}
\end{equation}
This integral has  a complicated 
analytical solution in terms of the
Whittaker function  $M_{\kappa,\mu}\left(z\right)$, see \cite{NIST2010}.
Figure \ref{wall_all} reports  all of the galaxies  of SDSS DR12
and also the theoretical curve. 

\begin{figure}
\begin{center}
\includegraphics[width=10cm]{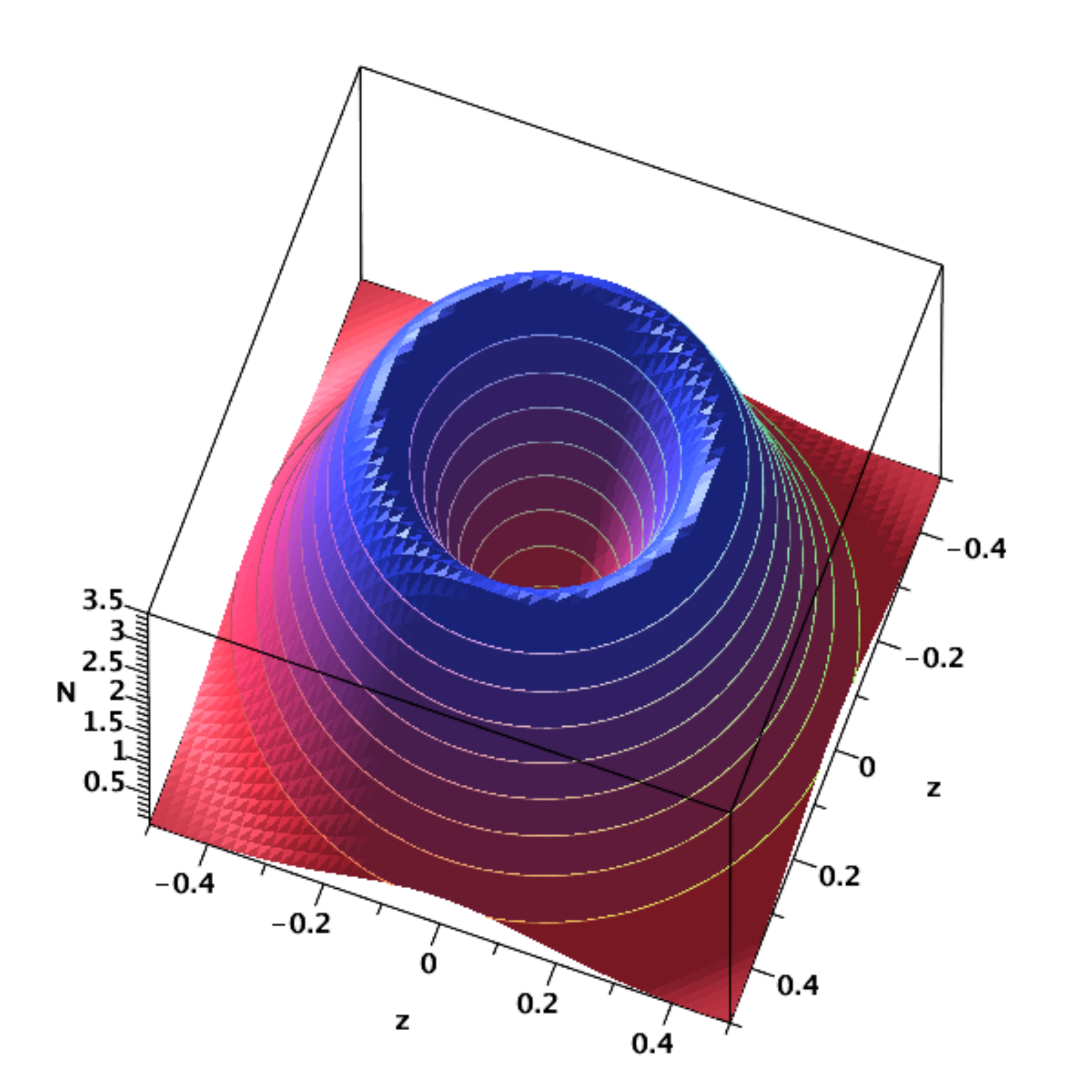}
\end{center}
\caption
{
Theoretical surface/contour of the number of all the galaxies 
divided  by 1000000
as a function of  redshift.
Parameters  as in Figure \ref{wall_all}.
}
          \label{wall_hole}%
    \end{figure}

A theoretical surface/contour  of the Great Wall is 
displayed in Figure \ref{wall_hole} 

\section{Conclusions}

{\bf $\Lambda$CDM cosmology}

In this paper, we use  the framework of $\Lambda$CDM cosmology
with parameters 
$H_0 = 69.81 \h0units$, $\om=0.239$  and  $\ola=0.651$.
A  relationship for the luminosity 
distance is  derived using the method of  the minimax approximation
when $p=6$ and $q=2$, see equation (\ref{dlz}).
The inverse relationship, the  redshift as function of
the luminosity function is derived in 
equation (\ref{zinverse}).

{\bf The Great Wall}

The enhancement in the number of galaxies  as a function of the 
redshift  
for the  SDSS Photometric Catalogue DR 12, 
which is  at  $z=0.383$,  
is here modeled by the theoretical equation 
(\ref{nzwall}) that is derived in the framework
of the Schechter LF for galaxies and  
the $\Lambda$CDM cosmology .
Figure \ref{wall_all} reports the observed  maximum in the number
of galaxies and also the  theoretical curve.
These results  are in agreement  with 
a catalog of photometric redshift of $\approx$ 3000000 SDSS DR8 galaxies
made by \cite{Paul2018} : their Figure 9 bottom reports that the count 
of elliptical galaxies  has a peak  at $z \approx  0.37$
when the spirals galaxies conversely peaks 
at $z \approx  0.08$.

\section*{Acknowledgments}

This research has made use of the VizieR catalogue access tool, CDS,
Strasbourg, France.


\begin{thebibliography}{10}
\expandafter\ifx\csname url\endcsname\relax
  \def\url#1{{\tt #1}}\fi
\expandafter\ifx\csname urlprefix\endcsname\relax\def\urlprefix{URL }\fi
\providecommand{\eprint}[2][]{\url{#2}}

\bibitem{Geller1989}
{Geller} M~J and {Huchra} J~P 1989 {Mapping the universe} {\em Science\/} {\bf
  246}, 897

\bibitem{deLapparent1989}
{de Lapparent} V, {Geller} M~J and {Huchra} J~P 1989 {The luminosity function
  for the CfA redshift survey slices} {\em \apj\/} {\bf 343}, 1

\bibitem{Ramella1992}
{Ramella} M, {Geller} M~J and {Huchra} J~P 1992 {The distribution of galaxies
  within the 'Great Wall'} {\em \apj\/} {\bf 384}, 396

\bibitem{Deng2007}
{Deng} X~F, {He} J~Z, {He} C~G and et~al 2007 {The Sloan Great Wall from the
  SDSS Data Release 4} {\em Acta Physica Polonica B\/} {\bf 38}, 219

\bibitem{Einasto2010}
{Einasto} M, {Tago} E, {Saar} E and et~al 2010 {The Sloan great wall. Rich
  clusters} {\em \aap\/} {\bf 522} A92 (\textit{Preprint} \eprint{1007.4492})

\bibitem{Einasto2011}
{Einasto} M, {Liivam{\"a}gi} L~J, {Tempel} E and et~al 2011 {The Sloan Great
  Wall. Morphology and Galaxy Content} {\em \apj\/} {\bf 736} 51
  (\textit{Preprint} \eprint{1105.1632})

\bibitem{Einasto2017}
{Einasto} M, {Lietzen} H, {Gramann} M and et~al 2017 {BOSS Great Wall:
  morphology, luminosity, and mass} {\em \aap\/} {\bf 603} A5
  (\textit{Preprint} \eprint{1703.08444})

\bibitem{dellAntonio1996a}
{dell'Antonio} I~P, {Bothun} G~D and {Geller} M~J 1996 {Peculiar Velocities for
  Galaxies in the Great Wall.I.The Data} {\em \aj\/} {\bf 112}, 1759

\bibitem{dellAntonio1996b}
{dell'Antonio} I~P, {Geller} M~J and {Bothun} G~D 1996 {Peculiar Velocities for
  Galaxies in the Great Wall.II.Analysis} {\em \aj\/} {\bf 112}, 1780

\bibitem{Zeldovich1970}
{Zel'dovich } Y~B 1970 {Gravitational instability: An approximate theory for
  large density perturbations.} {\em \aap\/} {\bf 5}, 84

\bibitem{Shandarin2009}
{Shandarin} S~F 2009 {The origin of `Great Walls'} {\em \jcap\/} {\bf 2} 031
  (\textit{Preprint} \eprint{0812.4771})

\bibitem{Shandarin2011}
{Shandarin} S~F 2011 {The multi-stream flows and the dynamics of the cosmic
  web} {\em \jcap\/} {\bf 5} 015 (\textit{Preprint} \eprint{1011.1924})

\bibitem{Valkenburg2012}
{Valkenburg} W and {Bj{\ae}lde} O~E 2012 {Cosmology when living near the Great
  Attractor} {\em \mnras\/} {\bf 424}, 495 (\textit{Preprint}
  \eprint{1203.4567})

\bibitem{Zaninetti2016a}
{Zaninetti} L 2016 Pade approximant and minimax rational approximation in
  standard cosmology {\em Galaxies\/} {\bf 4}(1), 4 ISSN 2075-4434
  \urlprefix\url{http://www.mdpi.com/2075-4434/4/1/4}

\bibitem{Suzuki2012}
{Suzuki} N, {Rubin} D, {Lidman} C, {Aldering} G, {Amanullah} R, {Barbary} K and
  {Barrientos} L~F 2012 {The Hubble Space Telescope Cluster Supernova Survey.
  V. Improving the Dark-energy Constraints above z greater than 1 and Building
  an Early-type-hosted Supernova Sample} {\em \apj\/} {\bf 746} 85

\bibitem{schechter}
{Schechter} P 1976 {An analytic expression for the luminosity function for
  galaxies.} {\em \apj\/} {\bf 203}, 297

\bibitem{Alam2015}
{Alam} S, {Albareti} F~D, {Allende Prieto} C and et~al 2015 {The Eleventh and
  Twelfth Data Releases of the Sloan Digital Sky Survey: Final Data from
  SDSS-III} {\em \apjs\/} {\bf 219} 12 (\textit{Preprint} \eprint{1501.00963})

\bibitem{NIST2010}
Olver F~W~J~e, Lozier D~W~e, Boisvert R~F~e and Clark C~W~e 2010 {\em {NIST
  handbook of mathematical functions.}\/} (Cambridge: {Cambridge University
  Press. })

\bibitem{Paul2018}
{Paul} N, {Virag} N and {Shamir} L 2018 A catalog of photometric redshift and
  the distribution of broad galaxy morphologies {\em Galaxies\/} {\bf 6}(2)

\end{thebibliography}
\providecommand{\newblock}{}

\end{document}